DEC-TR-301

# A Quantitative Measure of Fairness and Discrimination for Resource Allocation in Shared Computer System

Rajendra K. Jain, Dah-Ming W. Chiu, William R. Hawe

Eastern Research Lab

# A Quantitative Measure of Fairness and Discrimination for Resource Allocation in Shared Computer Systems

Rajendra K. Jain, Dah-Ming W. Chiu, William R. Hawe
Digital Equipment Corporation
77 Reed Road (HLO2-3/N03)
Hudson, MA 01749

## ABSTRACT

Fairness is an important performance criterion in all resource allocation schemes, including those in distributed computer systems. However, it is often specified only qualitatively. The quantitative measures proposed in the literature are either too specific to a particular application, or suffer from some undesirable characteristics. In this paper, we have introduced a quantitative measure called Index of Fairness. The index is applicable to any resource sharing or allocation problem. It is independent of the amount of the resource. The fairness index always lies between 0 and 1. This boundedness aids intuitive understanding of the fairness index. For example, a distribution algorithm with a fairness of 0.10 means that it is unfair to 90% of the users. Also, the discrimination index can be defined as 1 - fairness index.

The development of the proposed index is presented in this paper. A number of examples (taken particularly from computer networks) are illustrated in various contexts.

Revised September 26, 1984
PLEASE DO NOT MAKE COPIES

# 1. Introduction

Fairness is an important consideration in most performance studies. Particularly in distributed systems, where a set of resources is to be shared by a number of users, fair allocation is important. However, quantitative measures for fairness are not well known. Most studies on fairness tend to be either qualitative or too specific to a particular application. Any scheme resulting in uneven allocation of throughput is termed unfair. Jaffe wrote [6, 7] that a flow control scheme is fair if *"each user's throughput is at least as large as that of all other users which have the same bottleneck."* Gerla and Staskauskos [5] achieved this by minimizing an objective function originally proposed by Gallager and Golestaani [3]. The objective function includes a penalty for uneven allocation. Bharath-Kumar and Jaffe [2] state that any algorithm which allocates zero throughput to any user is unfair; otherwise, it is fair. These papers divide algorithms into two categories: fair or unfair. It is not possible to measure the fairness of a particular algorithm. Marsan and Gerla [8] present a fairness measure which which responds to this problem. They define fairness measure $F$ as:

$$min_{i,j}\left(\frac{p_i}{p_j}\right)$$

where $p_i$ and $p_j$ are powers (ratio of throughput and round trip delay) achieved by user $i$ and $j$ respectively. A channel access protocol is called optimally fair if $F=1$. Sauve, Wong, and Field [11] contend that fairness of a network is measured in terms of variance of network delays. The fairness definition is revised to square of coefficient of variation of delay by Wong and Lam [12], and then to a weighted measure of variance of delays [13].

It is clear from the above literature that fairness implies equal allocation of resources, although there is little agreement among researchers as to what should be equalized. In computer networks, some want equal delay, others want equal throughput, and yet others want equal power for all users sharing a resource. In other applications also, disagreements may exist about the proper allocation metric.

In this paper we divide the question of fairness into two parts. First, one needs to select an appropriate allocation metric which depends upon the application and desires of users. We leave open the selection of the metric. It could be any one, or a combination of different metrics used in a specific application. Second, one needs to define a formula which gives a quantitative value to the fairness of the allocation. It is this second aspect of fairness debate to which this paper is devoted. We present a formula which measures "equality" of resource allocation and, if not equal, it tells how far the allocation is from equality.

This paper is organized as follows: In the next section we present desired properties of a fairness index and discuss shortcomings of indices presented in other literature. Then we describe our proposed definition of "fairness index" and show its relationship to other indices. A short discussion on selection of the allocation metric is presented in section 5. The interpretation of the proposed index in terms of individual perception of fairness is described and a few properties of the index are discussed. Finally, we show its application to computer networks.

## 2. Desired Properties of a Fairness Index

In our search for a good index of fairness we set a number of goals. We found that the indices proposed in the literature did not satisfy some of these. If a system allocates resources to $n$ contending users such that the $i^{th}$ user receives an allocation

$x_i$, then the fairness measures proposed in the literature are:

1. $Variance = \dfrac{1}{n-1} \sum_{i=1}^{n} (x_i - \mu)^2$  where mean $\mu = \dfrac{1}{n} \sum_{i=1}^{n} x_i$

2. $Coefficient\ of\ Variation = \dfrac{Variance}{Mean}$

3. $Min-max\ ratio = \dfrac{min_j \{x_i\}}{max_j \{x_j\}} = min_{i,j} \left\{ \dfrac{x_i}{x_j} \right\}$

For example, if an algorithm results in three users receiving allocations of 1, 3, and 5 in dollars, then the fairness of this algorithm is respectively:

1. $Variance = 4,\quad \mu = 3$

2. $Coefficient\ of\ Variation = 1.33$

3. $Min-max\ ratio = \dfrac{1}{5}$

We feel the desired properties should be the following:

1. *Population Size Independence:* The index should be applicable to any number of users, finite or infinite. In particular, it should be applicable to two users sharing a resource. This requirement is satisfied by the above three indices.

2. *Scale and Metric Independence:* The index should be independent of scale, i.e., the unit of measurement should not matter. For example, the above algorithm allocates 100, 300, and 500 pennies respectively to the three users. The unfairness measured by variance now is 40,000 which seems 10,000 times as bad as before, even though we have not changed the algorithm. Thus, this property rules out the use of variance as a fairness index. Notice that coefficient of variation and min-max ratio are unaffected by scale.

3. *Boundedness:* The index should be bounded between 0 and 1, i.e., a totally fair system should have a fairness of 1 and a totally unfair system should have an index of 0. Thus, fairness can be expressed as a percentage. For example, a scheme with a fairness of 0.1 could be shown to be fair to 10% of the users and unfair to 90%. This helps in intuitive understanding of the index. Also, this allows us to define a discrimination index as 1 - fairness index.

The coefficient of variation (COV) can be anywhere between 0 and infinity. It is not easy to interpret what level of fairness is implied by a COV of 359 for instance. The min-max ratio is better in this respect in that it does satisfy this requirement.

4. *Continuity:* The index should be continuous. Any slight change in allocation should show up in the fairness index. In the above example, if the three users received 1, 4, and 5 dollars respectively, the fairness should obviously be different, yet it is not reflected in the min-max ratio which remains at 1/5.

Thus, we see that none of the known indices satisfies all requirements. This led us to the proposition of a new index.

### 3. Fairness Index: Proposed Definition

Again, if a system allocates resources to $n$ contending users, such that the $i^{th}$ user receives an allocation $x_i$, then we propose the following index for the system:

$$f(x) = \frac{\left[\sum_{i=1}^{n} x_i\right]^2}{n \sum_{i=1}^{n} x_i^2} \quad x_i \geq 0 \quad (1)$$

- 5 -

This index measures the "equality" of user allocation $x$. If all users get the same amount, i.e., $x_i$'s are all equal, then the fairness index is 1, and the system is 100% fair. As the disparity increases, fairness decreases and a scheme which favors only a selected few users has a fairness index near 0.

Notice that the proposed index satisfies all the requirements discussed before. It is dimensionless and independent of scale, it is bounded between 0 and 1, and it is continuous so that any slight change in $x_i$ changes the index.

The fairness index as defined here has a very intuitive interpretation as illustrated by the following example:

*Example 1:*

Suppose one is asked to distributed 20 dollars among 100 persons. Consider two ways to distribute the money:

Scheme A: Give 20 cents to each of the 100 persons.

In this case, $x_i = 0.2 \quad i = 1,2,...,100$

$$\text{Fairness Index} = f_A(x) = \frac{\left[\sum x_i\right]^2}{n \sum x_i^2} = 1.0$$

Scheme A is totally fair.

Scheme B: Depending upon age, sex, color, race, or any other discrimination criteria, choose 10 persons and give them 2 dollars each. Other 90 persons get no money.

In this case,

$$x_i = \begin{cases} 2 & i = 1, 2, 3, \ldots 10 \\ 0 & i = 11, 12, \ldots, 100 \end{cases}$$

$$\text{Fairness Index } f_B(x) = \frac{\left[\sum x_i\right]^2}{n \sum x_i^2} = 0.10$$

Scheme B is only 10% fair.

If in scheme B, we had selected 20 of the 100 people to receive $1 each, the fairness would have come out to 20%. This intuitive interpretation of fairness index is true in general. That is, given $m$ dollars to be distributed among $n$ people, if we select $k$ people, give them $m/k$ dollars each and give nothing to the rest, the fairness index would be $k/n$, or the fraction of people favored. Notice that the index does not depend upon the amount of the resource ($m$ dollars) or the population size ($n$ people). In the above example, the population size could have been 2 and the fairness could still be quantified.

The boundedness between 0 and 1 makes it easy to define a discrimination index. A scheme which is 10% fair is 90% unfair. Hence,

$$\text{Discrimination Index} = 1 - \text{Fairness index} = 1 - \frac{\left[\sum_{i=1}^{n} x_i\right]^2}{n \sum_{i=1}^{n} x_i^2}$$

*Example 2:*

A single resource is to be shared/divided between two consumers. The distribution algorithm gives $x$ to one consumer and $1-x$ resource to the other.

$$\text{fairness } f(x) = \frac{x + 1 - x}{2\left[x^2 + (1-x)^2\right]}$$

$$= \frac{0.5}{x^2 + (1-x)^2}$$

Figure 1 shows the fairness as a function of $x$. When both consumers get equal *resource*, the fairness is 1, and when one of the two is starved, the fairness is one-half.

Fairness does not necessarily mean equal distribution of resources. In some cases, it is justifiable to give more resources to some consumers than others. In such cases, the fairness index (equation 3) still applies. It is just that the appropriate performance metric in such cases would be the ratio of the resource allocated and the "right" for allocation. This is explained in the next example.

*Example 3:*

A single resource is to be distributed to two consumers. Consumer A pays twice as much as consumer B. The distribution algorithm allocates $x$ amount of resource to consumer A and $1-x$ to consumer B. Hence,

$$Fairness\ f(x) = \frac{\left[\frac{x}{2} + (1-x)\right]^2}{2\left[\left(\frac{x}{2}\right)^2 + (1-x)^2\right]}$$

Figure 2 shows the fairness as a function of $x$. When consumer A gets two-thirds (twice as much as consumer B), the fairness is 1.

## 4. Relationship to Other Fairness Definitions

If the allocations $x_i$ follow a certain random distribution, then the fairness index as defined here can be expressed as a function of the first two moments of $x_i$'s as shown below.

$$f(x) = \frac{\left[\sum x_i\right]^2}{n \sum x_i^2} = \frac{\left[\frac{1}{n}\sum x_i\right]^2}{\frac{1}{n}\sum x_i^2} = \frac{b_1^2}{b_2} \qquad (2)$$

where $b_1$ is the first moment of $x$ (also called mean) and $b_2$ is the second moment of $x$ about the origin.

The fairness definitions proposed in other literature are generally "variances" [11], "squared coefficient of variation" [13], or a "weighted-squared coefficient of variation" [13] of a pre-decided performance metric (usually path delay). Since coefficient of variation can also be calculated from the first and second moments, it is interesting to see how the fairness index proposed here is related to it. If $E(x)$ denotes the expected value of $x$,

$$Variance = E(x^2) - E^2(x)$$
$$= b_2 - b_1^2$$

$$and\ Coefficient\ of\ Variation\ (COV) = \frac{Standard\ deviation}{mean} = \frac{\left(b_2 - b_1^2\right)^{\frac{1}{2}}}{b_1}$$

$$fairness\ index = \frac{b_1^2}{b_2} = \frac{b_1^2}{b_1^2 + b_2 - b_1^2} = \frac{1}{1 + \frac{b_2 - b_1^2}{b_1^2}} = \frac{1}{1 + COV^2} \qquad (3)$$

Equation 3 shows the relationship between the fairness index proposed here and the coefficient of variation to be a simple transformation. However, it is an important transformation because it gives the fairness index all the desirable characteristics discussed below.

First, the proposed index is dimensionless and independent of the scale of the allocation metric.

Secondly, the relationship between coefficient of variation and fairness is an inverse one. That is, the system is totally fair if the coefficient of variation is zero. As the COV increases, fairness goes down. Transformation 3 removes this negativity. Thus, as the fairness increases the fairness index grows higher. This helps intuitive understanding of the results.

Thirdly, the coefficient of variation is an unbounded quantity. Thus, the fairness of a system (in previous literature) could be in hundreds or thousands. Transformation 3 makes fairness a bounded quantity between 0 and 1. A totally fair system has a fairness of 1 and a totally unfair system has a fairness of 0. This allows fairness to be expressed as a percentage; a system with a fairness index of 0.1 is only 10% fair. As shown in the previous section, "percent fairness" is not only a matter of convenience, it is also a meaningful interpretation. A scheme which allocates all resources equally to only 10% of the user population turns out to have a fairness index of 0.1. The squared coefficient of variation in that case would have been 9, which is difficult to interpret.

Equation 2 also brings up another interesting point. Since other ratios of moments are called "coefficients", for example:

$$Coefficient\ of\ Skewness = \frac{Third\ moment\ about\ the\ mean}{(Second\ moment\ about\ the\ mean)^{3/2}}$$

$$Coefficient\ of\ Kurtorsis = \frac{Fourth\ moment\ about\ the\ mean}{(Second\ moment\ about\ the\ mean)^{2}}$$

$$\text{Coefficient of Variation} = \frac{(\text{Second moment about the mean})^{1/2}}{\text{mean}}$$

we can define one more coefficient:

$$\text{Coefficient of fairness} = \frac{(\text{First moment about the origin})^2}{\text{Second moment about the origin}}$$

However, this coefficient is meaningful only for non-negative distributions (such as constant, exponential, uniform, log-normal, etc.) and is not meaningful for distributions that allow negative values (such as normal).

Coefficient of fairness for some of the distributions are shown in table 1.

## 5. Selection of the Allocation Metric

Notice that the fairness index is denoted as $f(x)$. The parameter $x$ signifies that the fairness based on another allocation metric would be different. The choice of the metric depends upon the application. Examples of metrics appropriate for different applications in networking are listed below.

1. Response Time: For interactive traffic, response time is generally the key performance indicator. If users going longer distances pay more, they may demand the same response time as those travelling short distances. In this case, the fairness should be based on equality of response time.

2. Response Time Per Hop: If all users are paying equally regardless of the distance travelled (as is generally the case in intra-corporation networks when all expenses are paid by a central network management), users going longer distances can expect longer response times than those going nearby. In such cases, fairness could be based on equality of "response time per hop" or "response time per mile."

3. Throughput: For file traffic, the key performance parameter is throughput. Hence, the fairness could be based on equality of throughput.

4. Throughput Times Hops: This is based on the argument that long distance users should expect to get less throughput, particularly if they are not paying any more than short distance users.

5. Power: If the traffic consists of both file and terminal traffic, the fairness may be based on the ratio of throughput and response time, i.e., power, or other functions of the two.

6. Fraction of Demand: In a system where users make unequal demands for resources, one may want to measure fairness by closeness of the allocations to respective demands. One way to do this is to define $x_i$ as follows:

$$x_i = \begin{cases} \dfrac{a_i}{d_i} & \text{if } a_i < d_i \\ 1 & \text{Otherwise} \end{cases}$$

Here, $d_i$ is the demand of $i^{th}$ user and $a_i$ is the corresponding allocation. Allocating more than the demand does not make the user any happier.

The fairness definition proposed here can be applied to any metric; therefore, we have not chosen any particular metric or application. The discussion applies as much to sociology problems of fairness of income distribution as it does to fairness of waiting times of customers in a grocery store, or to computer networks. An important consideration in the choice of performance metric is that it must be not-negative for all users; all $x_i$'s must be non-negative. In most resource allocation problems, this is usually the case; the minimum a user can get is no resource at all. (The extension of this concept to negative $x_i$'s is currently under investigation.)

## 6. User Perception of Fairness

One way to rewrite the proposed formula for fairness is:

$$f(x) = \frac{1}{n} \sum_{i=1}^{n} \frac{x_i}{x_f}$$

where $x_f$ is the *fair allocation mark* computed as follows:

$$x_f = \frac{\sum_{i=1}^{n} x_i^2}{\sum_{i=1}^{n} x_i} = \frac{b_2}{b_1}$$

Thus, each user compares his allocation $x_i$ with the amount $x_f$, and perceives the algorithm as fair or unfair depending upon whether his allocation $x_i$ is more or less that $x_f$. For the $i^{th}$ user, the algorithm is only $x_i/x_f$ fair. The overall fairness is the average of *perceived fairness* of all $n$ users.

In example 1 (of $20 distributed among 100 users), using scheme B,

$$\text{Fair Allocation} = \frac{\sum_{i=1}^{n} x_i^2}{\sum_{i=1}^{n} x_i} = \frac{\sum_{i=1}^{10} 2^2 + \sum_{i=11}^{100} 0^2}{\sum_{i=1}^{10} 2 + \sum_{i=11}^{100} 0} = 2$$

Thus, the 10 users who received $2 perceive the scheme to be 100% fair, and the 90 users who didn't receive anything perceive the scheme to be 0% fair. The overall fairness is 10%. All users with $x_i > x_f$ are said to be *favored* users and those with $x_i < x_f$ are *discriminated* users.

The *perceived discrimination* of the $i^{th}$ user is $(x_f - x_i)/x_f$ and the discrimination index is the average of perceived discrimination:

$$\text{Discrimination Index} = \frac{1}{n} \sum_{i=1}^{n} \frac{x_f - x_i}{x_f}$$

Among $n$ users, not all users can be discriminated; therefore,

$$f(x) \geq \frac{1}{n}$$

Notice that $f(x) = \mu/x_f$ where $\mu$ is the average of $x_i$'s. Hence $x_f = \mu/f$. The fair allocation mark $x_f$ is always more than or equal to the average since $f \leq 1$.

## 7. Properties of the Proposed Fairness Function

In this section, we present a few simple results regarding behavior of the fairness index as allocations are changed. The proofs of the theorems are in Appendix A.

1. Resource Exchange: If we take a small amount of resource from a user and give it to another user, the new allocation should be more fair if the receiver has less resource than the giver and vice versa. The proposed index does satisfy this property as the following theorem states:

*Theorem 1:* If $\Delta x$ resource is taken away from a user $k$ and given to another user $j$, then the fairness

1. increases iff $\Delta x < x_k - x_j$
2. remains same iff $\Delta x = x_k - x_j$
3. decreases iff $\Delta x > x_k - x_j$

2. **Additional Allocation:** If each user is given an additional amount of resource, we would expect individual perception of fairness to rise and the overall fairness to go up. On the other hand, if only a single user is given additional resources, others may become dissatisfied if the user is already a 'favored user.' These properties of the fairness index are stated by the following theorem:

*Theorem 2a:* If each user is given an additional amount $c$ of the resource, their individual perception of fairness increases and so does the overall fairness index:
$$f(x_1 + c, + x_2 + c, \ldots, x_n + c) \geq f(x_1, x_2, \ldots, x_n)$$

*Theorm 2b:* If a single user $j$ is given a small additional allocation $\Delta x_j$ without changing other allocations, the new allocation is more (less) fair than before iff $j$ is a discriminated (favored) user, i.e.,

$$f(x_1, x_2, \ldots, x_{j-1}, x_j + \Delta x_j, x_{j+1}, \ldots, x_n) > f(x_1, x_2, \ldots, x_{j-1}, x_j, x_{j+1}, \ldots, x_n) \text{ if } x_j < x_f$$

$$f(x_1, x_2, \ldots, x_{j-1}, x_j + \Delta x_j, x_{j+1}, \ldots, x_n) \, f(x_1, x_2, \ldots, x_{j-1}, x_j, x_{j+1}, \ldots, x_n) \text{ if } x_j > x_f$$

3. **Maximization:** The fairness index has a bell-shaped behavior curve with respect to each individual allocation. As a particular user's allocation is increased from zero, the fairness first increases. It then reaches a maximum. Any further allocation to that user results in other users perceiving unfairness and the overall fairness decreases.

*Theorem 3:* If we vary single user $j$'s allocation, (while not affecting other users' allocations), the fairness reaches maximum when

$$x_j = x_{f, n-1}$$

where $x_{f, n-1}$ is the fair allocation mark for the remaining $n-1$ users, i.e.,

$$x_{f, n-1} = \frac{\sum_{i \neq j} x_i^2}{\sum_{i \neq j} x_i}$$

4. Bounded Allocations: If there are no limits on allocation, the worst case of fairness can be near zero. By putting upper and lower bounds on allocation, so that no user can get less than a certain amount or more than a certain amount, we can guarantee a minimum level of fairness.

*Theorem 4:* If $x_{min} \leq x_i \leq x_{max}$ for all $i$ and $x_{min} > 0$, then the fairness is guaranteed to be above a certain lower bound, and

a. The minimum fairness occurs when a fraction $\gamma$ of the users (i.e., $\gamma n$ users in all) receive $x_{min}$ and the remaining receive $x_{max}$. Here

$$\gamma = \frac{K}{K+1}$$

$$\text{where } K = \frac{x_{max}}{x_{min}}$$

b. fairness

$$f \geq \frac{4K}{(K+1)^2}$$

## 8. Fairness of Computer Networks

Fairness is an important criterion in designing computer networks. There are many factors which influence the performance perceived by the network users. A key factor is the flow control scheme. In this section, we provide a simple illustration of how the flow control scheme affects fairness. In particular, we consider networks with window flow control scheme [4]. In such a network, each user sets a limit on the maximum number of outstanding packets (packets not yet acknowledged). This limit is called window size. IBM's SNA network, for example, uses a scheme in which the users vary the window size between a specified maximum and minimum [1].

It is known that a virtual circuit with a window size of $C$ can be modelled with a closed queueing network of population size $C$ [9]. For a h-hop path, the queueing network has $h + 1$ nodes. Suppose there are $n$ users (virtual circuits) sharing the same physical path and let

$$c_i = \text{Window size of } i^{th} \text{ user}$$

$$\text{and } C = \sum_{i=1}^{n} c_i = \text{Sum of all window sizes.}$$

We assume that packet lengths are distributed such that the service time at each node is exponentially distributed with a mean of 1. This system can be modelled by a closed queueing system of $h + 1$ nodes and a population size of $C$ as shown in figure 3. This is a rather simplistic model; however it has been chosen since it helps in understanding the effect of window size on fairness. Using mean value analysis [10], the following performance metrics can be derived:

$$\text{Throughput of } i^{th} \text{ user} = T_i = \frac{c_i}{h + \sum_{i=1}^{n} c_i}$$

$$\text{Response time for } i^{th} \text{ user} = R_i = h + \sum_{i=1}^{n} c_i$$

$$\text{Power of } i^{th} \text{ user} = P_i = \frac{\text{Throughput of } i^{th} \text{ user}}{\text{Response time for } i^{th} \text{ user}} = \frac{c_i}{(h + \sum_{i=1}^{n} c_i)^2}$$

Using these formulas, a number of statements can be made regarding fairness:

1.  Fairness based on Response Time:

    If the goal of the system is to provide the same mean response time to all users, then the fairness index would be

    $$\text{fairness (response time)} = \frac{\left\{\sum_{i=1}^{n} R_i\right\}^2}{n \sum_{i=1}^{n} R_i^2} = 1.0$$

    i.e., the window size does not matter (for this simple example). As long as the users are travelling the same number of hops, they will have equal response time regardless of individual window size.

2.  Fairness based on Throughput:

    If the goal is to provide the same throughput to all users, the fairness index would be

    $$\text{fairness (throughput)} = \frac{\left\{\sum_{i=1}^{n} T_i\right\}^2}{n\left\{\sum_{i=1}^{n} T_i^2\right\}}$$

$$= \frac{\left[\sum_{i=1}^{n} c_i\right]^2}{n\left[\sum_{i=1}^{n} c_i^2\right]} = \textit{fairness (window size)} \qquad (4)$$

i.e., the fairness based on throughput would be the same as that based on window size. To provide equal throughput to all users, they all must have the same window size.

3. Fairness based on Power:

If the goal of the flow control scheme is to provide equal power to all users, the fairness index would be

$$\textit{fairness (power)} = \frac{\left[\sum_{i=1}^{n} P_i\right]^2}{n \sum_{i=1}^{n} P_i^2} = \frac{\left[\sum_{i=1}^{n} c_i\right]^2}{n \sum_{i=1}^{n} c_i^2}$$

$$= \textit{Fairness (window size)} = \textit{Fairness (throughput)}$$

In this case, the same design consideration applies as discussed above for equal throughput.

4. Fairness of Variable Window Flow Control:

In variable window size networks, the users are allowed to vary the window size within a specified range. For example, in SNA [1]

Minimum window size = $h$ = number of hops

Maximum window size = $3h$ = 3 times number of hops

To understand the impact of variable window size on fairness, suppose the maximum window size $c_{max}$ is $K$ times the minimum $c_{min}$ ($K$ = 3 in SNA), and

- 19 -

suppose the goal is to measure fairness based on throughput or power. This is a case of bounded allocations, and theorem 4 tells us that the minimum fairness would occur when some users have their windows set at the minimum and the rest have their windows at the maximum. If $\gamma$ is the fraction of users with their window at minimum, then

$$Fairness\ (throughput) = fairness\ (window\ size)$$

$$= \frac{\left\{\gamma n c_{min} + (1-\gamma) n c_{max}\right\}^2}{n\left\{\gamma n c_{min}^2 + (1-\gamma) n c_{max}^2\right\}}$$

$$= \frac{\left\{\gamma + (1-\gamma) K\right\}^2}{\gamma + (1-\gamma) K^2}$$

$$Where,\ K = \frac{c_{max}}{c_{min}}$$

Figure 4 shows the graph of fairness versus $\gamma$ for various values of $K$. Theorem 4 and this figure display that, for a given $K$, the fairness is guaranteed to be above a certain minimum, i.e.,

$$fairness \geq \frac{4K}{(K+1)^2}$$

This is, in fact, the lower bound on the right hand side of equation 4. For example, in SNA ($K = 3$), the fairness is guaranteed to be at least 75%. As $K$ is increased, and the range of window sizes is increased, the lower bound of fairness decreases and the network can become unfair to many more users (see figure 5).

## 9. Other Fairness Functions

The proposed index of fairness is not the only function that satisfies the requirements listed in section 2. In fact, given an allocation $x = \{x_1, x_2, \ldots, x_n\}$, any function of the form

$$f(x) = \frac{\left[\frac{1}{n}\sum x_i\right]^r}{\frac{1}{n}\sum x_i^r} = \frac{(First\ Moment)^r}{r^{th}\ Moment} \quad r > 1$$

seems to satisfy all the requirements, including those of continuity, boundedness, and dimensionlessness. However, the previously proposed function is still a preferred measure of fairness because of its "linear behavior" with respect to the fraction of favored users in Example 1. If $m$ dollars are to be distributed among $n$ people and we favor $k$ people giving them $m/k$ dollars each and discriminate against $n-k$ people, then the above function would give

$$fairness = \left(\frac{k}{n}\right)^{r-1}$$

i.e., favoring 10% would result in a fairness index of $0.1^{r-1}$ and discrimination index of $1 - 0.1^{r-1}$. Obviously, $r=2$ seems to be the correct choice.

## 10. Summary

The key contribution of this paper is to breakdown the problem of fairness in two parts: selection of appropriate allocation metric, and quantification of equality. We have proposed a formula for quantifying the equality. This fairness function has many desirable properties which other formulas do not satisfy. The proposed fairness index is independent of scale of the allocation metric. It is bounded between 0 and 1 so that it can be meaningfully expressed as a percentage. Finally, it is continuous so that any change in allocation changes the fairness also.

The fairness function applies to any system with shared resources; therefore, the discussion is independent of any particular application. We have provided a number of examples from different areas to illustrate applications in different situations. In particular, we have derived a lower bound for the fairness of computer networks with window flow control.

# Appendix A: Proofs of Theorems

**Proof of Theorem 1 (Resource Exchange):**

$$f = \frac{\left(\sum_{i=1}^{n} x_i\right)^2}{n \sum_{i=1}^{n} x_i^2}$$

If $\Delta x$ is taken away from $nk$ and given to $x_j$, the new allocations for the two users are

$$x'_j = x_j + \Delta x$$

$$x'_k = x_k - \Delta x$$

The new fairness index value is

$$f' = \frac{\left(\sum_{i=1}^{n} x_i\right)^2}{n\left\{(x_j + \Delta x)^2 + (x_k - \Delta x)^2 + \sum_{i \neq j,k} x_i^2\right\}}$$

$$= \frac{\left(\sum_{i=1}^{n} x_i\right)^2}{n \sum_{i=1}^{n} x_i^2 + 2n\Delta x\left[\Delta x - (x_k - x_j)\right]}$$

Since $f$ and $f'$ differ only in the denominators,

$$f' < f \text{ iff } \Delta x - (x_k - x_j) > 0$$

[Q.E.D.]

**Proof of Theorem 2a (Additional Allocation):**

Let

$$a = \sum_{i=1}^{n} x_i \quad \text{and} \quad b = \sum_{i=1}^{n} x_i^2$$

so that

Since
$$f(x) = \frac{a^2}{nb}.$$

$$f \leq 1, \quad a^2 \leq nb \, \forall n.$$

$$f(x_1 + c, x_2 + c, \ldots, x_n + c) - f(x_1, x_2, \ldots, x_n) = \frac{\left[\sum_{i=1}^{n}(x_i + c)\right]^2}{n\sum_{i=1}^{n}(x_i + c)^2} - \frac{\left(\sum_{i=1}^{n} x_i\right)^2}{n\sum_{i=1}^{n} x_i^2}$$

$$= \frac{(a + nc)^2}{n(b + 2ca + nc^2)} - \frac{a^2}{nb}$$

$$= \frac{b(a + nc)^2 - a^2(b + 2ca + nc^2)}{nb(b + 2ca + nc^2)}$$

$$= \frac{c(nb - a^2)(2a + nc)}{nb(b + 2ca + nc^2)}$$

$$> 0 \qquad \forall c > 0$$

[Q.E.D.]

**Proof of Theorem 2b** (Additional Individual Allocation):

Using $a$ and $b$ as defined in the previous proof,

$$f(x_1, x_2, \ldots, x_{j-1}, x_j + \Delta x_j, x_{j+1}, \ldots, x_n) - f(x_1, x_2, \ldots, x_{j-1}, x_j, x_{j+1}, \ldots, x_n)$$

$$= \frac{\left[(x_j + \Delta x_j) + \sum_{i \neq j} x_i\right]^2}{n\left[(x_j + \Delta x_j)^2 + \sum_{i \neq j} x_i^2\right]} - \frac{\left(\sum_{i=1}^{n} x_i\right)^2}{n\sum_{i=1}^{n} x_i}$$

$$= \frac{(a+\Delta x_j)^2}{n[b+2x_j\Delta x_j+(\Delta x_j)^2]} - \frac{a^2}{nb}$$

$$\frac{b(a^2+2a\Delta x_j)-a^2(b+2x_j\Delta x_j)}{nb(b+2x_j\Delta x_j)} \quad \text{Ignoring } (\Delta x_j)^2$$

$$= \frac{(2ab-2a^2x_j)\Delta x_j}{nb(b+2x_j\Delta x_j)}$$

$$= \frac{2a^2(x_f-x_j)\Delta x_j}{nb(b+2x_j\Delta x_j)} \quad \text{since } x_f = \frac{b}{a}$$

$$> 0 \quad \text{Iff } x_j < x_f$$

[Q.E.D]

## Proof of Theorem 3 (Maximization):

Let

$$b_1 = \frac{1}{n}\sum_{i=1}^{n} x_i = \text{First moment about the origin}$$

$$b_2 = \frac{1}{n}\sum_{i=1}^{n} x_i^2 = \text{Second moment about the origin}$$

So that

$$f = \frac{b_1^2}{b_2}$$

and

$$\frac{\delta b_1}{\delta x_j} = b_1' = \frac{1}{n}$$

$$\frac{\delta b_2}{\delta x_j} = b_2' = \frac{2 x_j}{n}$$

Therefore,

$$\frac{\delta f}{\delta x_j} = f' = \frac{2 b_1 b_1' b_2 - b_1^2 b_2'}{b_2^2}$$

$$= \frac{2}{n} \frac{b_1 b_2 - b_1^2 x_j}{b_2^2}$$

$$= \frac{2}{n} \frac{b_1 (b_2 - b_1 x_j)}{b_2^2}$$

$$\frac{\delta f}{\delta x_j} = 0 \Rightarrow b_2 - b_1 x_j = 0$$

$$\sum_{i=1}^{n} x_i^2 - x_j \sum_{i=1}^{n} x_i = 0$$

or

$$\sum_{\substack{i=1 \\ i \neq j}}^{n} x_i^2 - x_j \sum_{i \neq j} x_i = 0$$

$$x_j = \frac{\sum_{i \neq j} x_i^2}{\sum_{i \neq j} x_i}$$

$$= x_{f, n-1}$$

Further

$$\frac{\delta^2 f}{\delta^2 x_j} = \frac{2}{n} \left[ \frac{b_2^2 (b_1' b_2 + b_1 b_2' - 2 b_1 b_1' x_j - b_1^2) - 2 b_2 b_2' (b_1 b_2 - b_1^2 x_j)}{b_2^4} \right]$$

$$= \frac{2}{n} \frac{b_2^2 b_1 (b_2' - b_1 - b_1' x_j)}{b_2^4}$$

Substituting $b_1' = 1/n$ and $b_2' = 2 x_i/n$

$$\frac{\partial^2 f}{\partial^2 x_j} = \frac{2}{n^2} \frac{b_2^2(b_2 + 2 b_1 x_j - 2 b_1 x_j - n b_1^2) - 4 b_2 x_j (b_1 b_2 - b_1^2 x_j)}{b_2^4}$$

Substituting $x_j = b_2 / b_1$

$$\frac{\partial^2 f}{\partial^2 x_j} = \frac{2}{n^2} \frac{b_2^2 (b_2 + 2 b_2 - 2 b_2 - 2 b_2 - n b_1^2) - 4 b_2 (b_2 - b_2)}{b_2^4}$$

$$= \frac{2}{n^2} \frac{b_2 - n b_1^2}{b_2^2}$$

$$= \frac{2}{n^2} \frac{b_1 x_j - n b_1^2}{b_2^2}$$

$$= \frac{2}{n^2} \frac{b_1}{b_2^2} \cdot (n b_1 - x_j)$$

$$= -\frac{2}{n^2} \cdot \frac{b_1}{b_2^2} \sum_{i \neq j} x_i$$

$$< 0 \quad provided \sum_{i \neq j} x_i \neq 0$$

Hence, $f$ is maximum at $x_j$

$$x_j = \frac{\sum_{i \neq j} x_i^2}{\sum_{i \neq j} x_i} = x_{f, n-1}$$

[Q.E.D.]

**Proof of Theorem 4** (Bounded Allocations):

The proof consists of two parts. In the first part, we prove by contradiction that minimum fairness occurs for an allocation such that $m$ of $n$ users receive $x_{min}$ and the remaining $n$-$m$ users receive $x_{max}$; all users receive either $x_{min}$ or $x_{max}$, nothing in-between. In the second part, we find the number $m$ such that the fairness is minimum.

Part 1: Suppose the minimum fairness occurs for an allocation $(x_1, x_2, \ldots, x_n)$ in which at least one user, say $j$'s allocation is neither at $x_{min}$ nor $x_{max}$. For the remaining $n$-$1$ users, we calculate the fair allocation mark $x_{f n-1}$. From theorem 2b we know that if $x_j < x_{f, n-1}$ (i.e., $x_j$ is being discriminated), the fairness can be further reduced by slowly taking away some of the allocation till $x_j = x_{min}$. On the other hand, if $x_j > x_{f, n-1}$ (i.e., $x_j$ is being favored), the fairness can again be reduced by giving a little more till $x_j = x_{max}$. If $x_j = x_{f, n-1}$, then setting $x_j = x_{min}$ or $x_j = x_{max}$ both will decrease fairness. We will set $x_j$ to the value the of the lower fairness. Thus, in all possible cases, we see that fairness can be further reduced by setting $x_j$ at one of the two bounds.

Part II: Suppose we allocate $m$ users $x_{max}$ and the remaining $n$-$m$ users $x_{min}$, i.e.,

$$x_i = \begin{cases} x_{max} & i = 1, 2, \ldots, m \\ x_{min} & i = m+1, m+2, \ldots, n \end{cases}$$

then fairness

$$f = \frac{\left\{n x_{max} + (n-m)x_{min}\right\}^2}{n\left\{n x_{max}^2 + (n-m)x_{min}^2\right\}}$$

Letting $m = \gamma n$

and $x_{max} = K x_{min}$

we get

$$f = \frac{\left\{\gamma + (1-\gamma)K\right\}^2}{\gamma + (1-\gamma)K^2}$$

The rest of the proof is similar to that discussed in section 8 under fairness of SNA. The minimum can be found by the usual procedure of setting

$$\frac{\delta f}{\delta \gamma} = 0 \quad \frac{2\{\gamma + (1-2\gamma)K\}(1-K)\{\gamma + (1-\gamma)K^2\} - (1-K^2)\{\gamma + (1-\gamma)K\}^2}{\{\gamma + (1-\gamma)K^2\}}$$

$$= \frac{\{\gamma + (1-\gamma)K\}(1-K)[2\{\gamma + (1-\gamma)K^2\} - (1+K)\{\gamma + (1-\gamma)K\}]}{\{\gamma + (1-\gamma)K^2\}^2}$$

$$= \frac{\{\gamma + (1-\gamma)K\}(1-K)[(K-1)\{K-\gamma(K+1)\}]}{\{\gamma + (1-\gamma)K^2\}^2}$$

which gives

$$\gamma = \frac{K}{K+1}$$

and

$$f = \frac{4K}{(K+1)^2}$$

[Q.E.D.]

# REFERENCES

1. Ahuja, W. Routing and Flow Control in System Network Architecture. *IBM Systems Journal*, 18, 2, (1979) 298 - 314.

2. Bharath-Kumar, K., and Jaffe, J.M. A New Approach to Performance-Oriented Flow Control. *IEEE Transactions on Communications*, COM-29, 4, (April, 1981), 427 - 435.

3. Gallager, R.G., and Golestaani, S.J. Flow Control and Routing Algorithms for Data Networks. *Proc. Intl. Conf. on Computer Comm.*, Atlanta, GA, (October 1980), 779 - 784.

4. Gerla, M., Kleinrick, L. Flow Control: A Comparative Survey, *IEEE Transactions on Communications*, COM-28, 4, (April 1980), 553-574.

5. Gerla, M., and Staskauskos, M. Fairness in Flow Controlled Networks. *Preceedings IEEE 1981 International Conference on Communications*, 3, Denver, CO, (June 14 - 18, 1981), 63.2.1 - 5.

6. Jaffe, J.M. A Decentralized, "Optimal", Multiple-User, Flow Control Algorithm. *Proc. Intl. Conf. on Computer Comm.*, Atlanta, GA, (October 1980), 839 - 844.

7. Jaffe, J.M. Bottleneck Flow Control. *IEEE Transaction on Communications*, COM-29, 7, (July 1981), 954 - 962.

8. Marsan, M.A., and Gerla, M. Fairness in Local Computing Networks. *Proceedings IEEE International Conference on Communication ICC82*, Philadelphia, PA, (June 13 - 17, 1982), 2F.41 - 6.

9. Reiser, M. A Queueing Network Analysis of Computer Communication Networks with Window Flow Control. *IEEE Transactions of Communications*, COM-27, 8, (August 1979), 1199 - 1209.

10. Reiser, M., and Lavenberg, S.S. Mean Value Analysis of Closed Multichain Queueing Networks. *JACM*, 27, 2 (April 1980), 466 - 470.

11. Sauve, J.P., Wong, J.W., and Field, J.A. On Fairness in Packet-Switching Networks. *Proceedings 21st IEEE Computer Society International Conference, CompCon 80*, Washington, D.C., (September 23 - 25, 1980), 466-470.

12  Wong, J.W., and Lam, S.S. Queueing Network Models of Packet Switching Networks, Part 1: Open Networks. *Performance Evaluation*, 2, (1982), 9 - 21.

13  Wong, J.W., Sauve, F.P., and Field J.A. A Study of Fairness in Packet-Switching Networks. *IEEE Transaction on Communications*, COM-30, 2, (February 1982), 346 - 353.

## Table 1: Coefficient of Fairness for Distributions

| Distribution | Density Function | First Moment $E(x)$ | Second Moment $E(x^2)$ | Coefficient of Fairness $f(x)$ |
|---|---|---|---|---|
| Constant | $p(x) = 1$ iff $x = a$ | $a$ | $a^2$ | 1 |
| Exponential | $p(x) = \lambda e^{\lambda x}$ | $1/\lambda$ | $\dfrac{2}{\lambda^2}$ | 0.5 |
| Erlang | $p(x) = \dfrac{(\lambda x)^{c-1} \lambda e^{\lambda x}}{(c-1)!}$ | $\dfrac{c}{\lambda}$ | $\dfrac{c(c+1)}{\lambda^2}$ | $\dfrac{c}{c+1}$ |
| Uniform | $p(x) = \dfrac{1}{b-a}$ iff $a \le b$ | $\dfrac{a+b}{2}$ | $\dfrac{a^2 + ab + b^2}{3}$ | $\dfrac{4(a^2 + ab + b^2)}{3(a+b)^2}$ |
| Lognormal | $p(x) = \dfrac{1}{x\sigma(2\pi)^{1/2}} \exp\left\{\dfrac{-[\ln(x/m)]^2}{2\sigma^2}\right\}$ | $m e^{1/2 \sigma^2}$ | $m^2 e^{2\sigma^2}$ | $e^{-\sigma^2}$ |

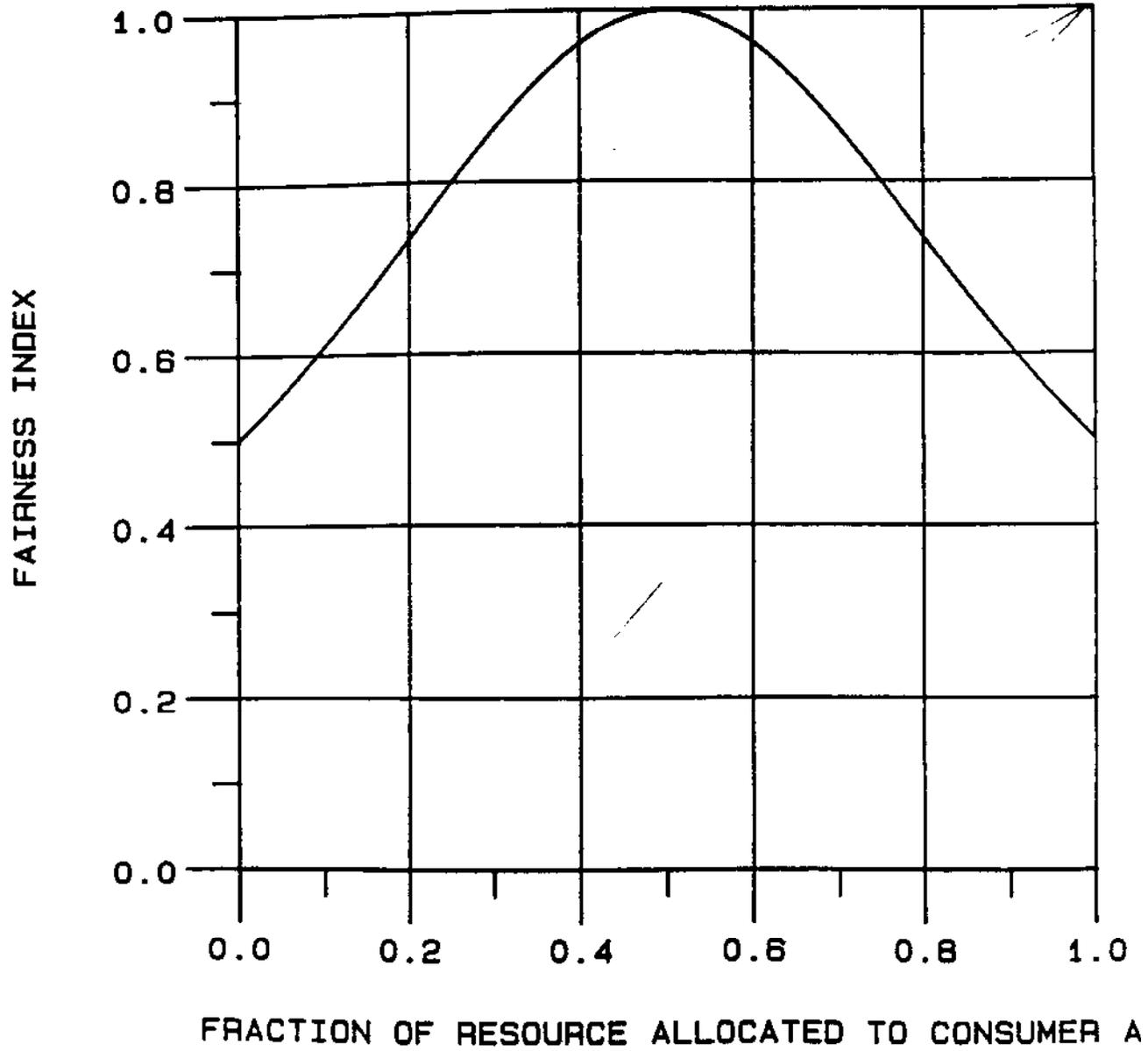

Figure 1: Resource allocation between two consumers A and B.

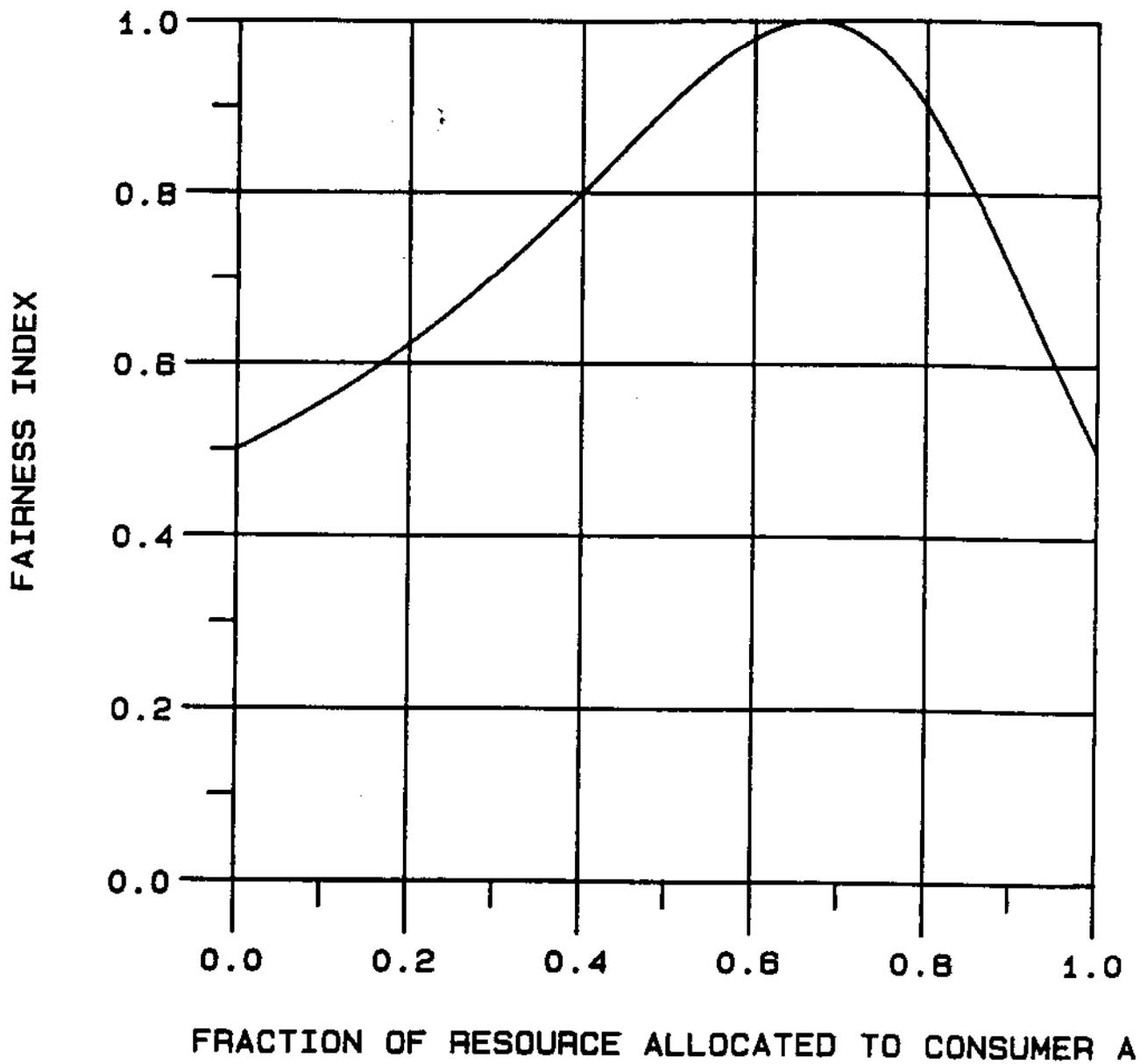

Figure 2: Consumer A deserves twice as much resource as consumer B.

**Figure 3:** A h-hop path shared by $n$ virtual circuits can be modelled by a closed queueing network of $h + 1$ servers and population size equal to sum of window sizes.

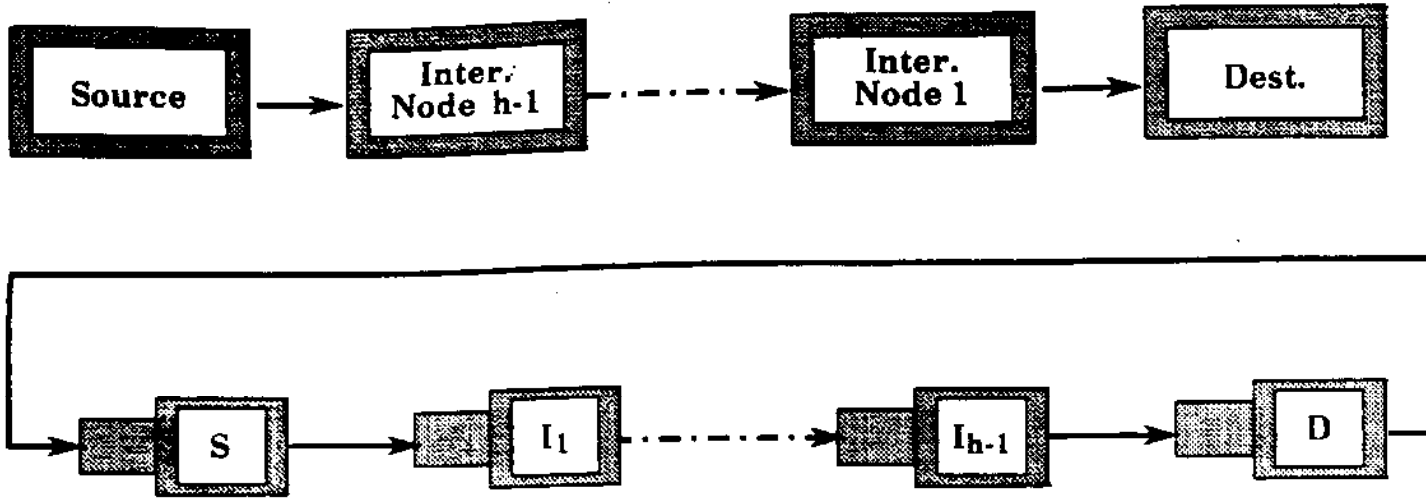

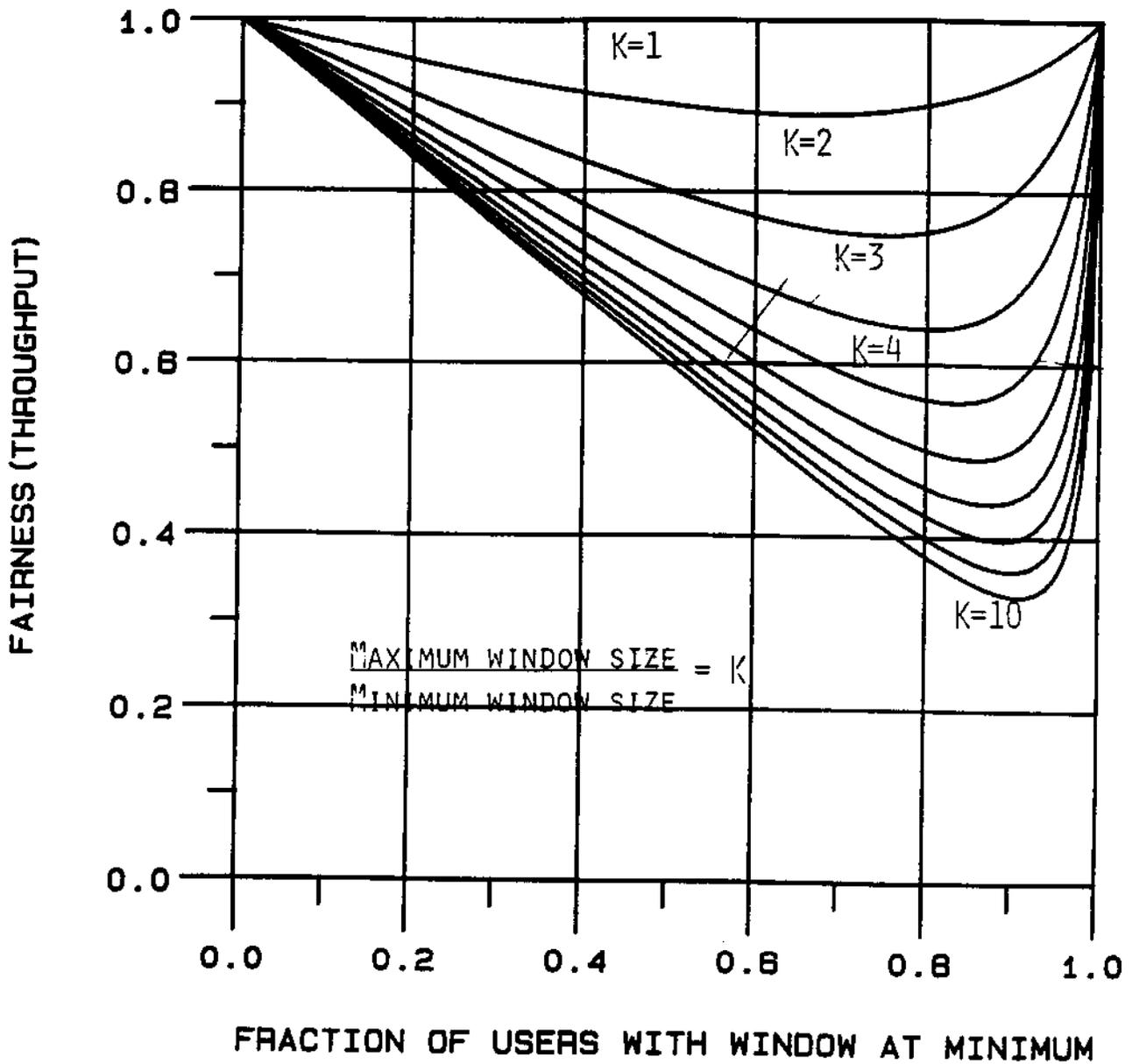

**Figure 4:** Fairness based on throughput for a network with variable window flow control.

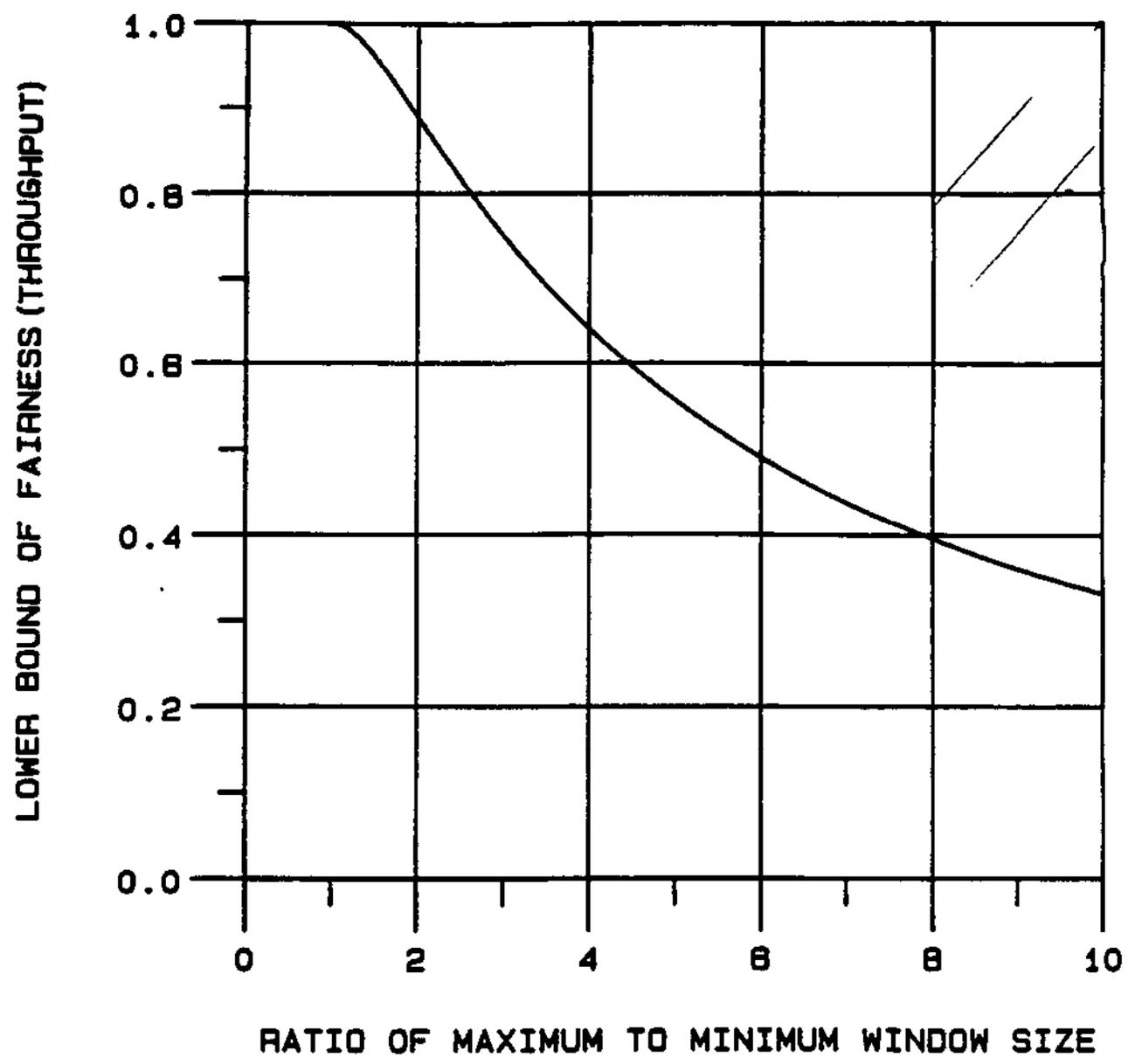

Figure 5: A variable window flow control scheme becomes more unfair as the range of allowed window size is widened.